%% file: root.tex
\documentclass[letterpaper, 10 pt, conference]{ieeeconf}  % Comment this line out if you need a4paper
\usepackage{xcolor}
\usepackage{graphicx}

\usepackage{cite}
\usepackage{amsmath,amssymb,amsfonts}
\usepackage{algorithm,algorithmic}
\usepackage{graphicx}
\usepackage{diagbox}
\usepackage{threeparttable}
\usepackage{textcomp}
\usepackage{booktabs}
\usepackage{epstopdf}
\usepackage{bm}
\usepackage{hyperref}
\usepackage{url}
\usepackage{multirow}
\usepackage{amsmath}
\usepackage{amssymb}

\usepackage{amsthm}
\usepackage{float}
\usepackage[justification=centering]{caption}

\IEEEoverridecommandlockouts                              % This command is only needed if 
\title{\LARGE \bf
Fast Constraint Screening for Multi-Interval Unit Commitment
}

\author{Xuan He$^{1}$, Jiayu Tian$^{2}$, Yufan Zhang$^{3}$, Honglin Wen$^{4}$ and Yize Chen$^{1}$% <-this % stops a space
\thanks{$^{1}$ Hong Kong University of Science and Technology (Guangzhou), {\tt\small xhe085@connect.hkust-gz.edu.cn, yizechen@ust.hk }}%
\thanks{$^{2}$ Sun Yat-Sen University,  {\tt\small tianjy23@mail2.sysu.edu.cn}}%
\thanks{$^{3}$ University of California, San Diego,  {\tt\small yuz254@ucsd.edu}}%
\thanks{$^{4}$ Shanghai Jiaotong University, {\tt\small linlin00@sjtu.edu.cn}}%
}%
\usepackage{cases}
\usepackage{amsmath,amsfonts}

\begin{document}

\maketitle
\thispagestyle{empty}
\pagestyle{empty}

\begin{abstract}
Power systems Unit Commitment (UC) problem determines the generator commitment schedule and dispatch decisions for power networks based on forecasted electricity demand. However, with the increasing penetration of renewables and stochastic demand behaviors, it becomes challenging to solve the large-scale, multi-interval UC problem in an efficient manner. The main objective of this paper is to propose a fast and reliable scheme to eliminate a set of redundant or inactive physical constraints in the high-dimensional, multi-interval, mixed-integer UC problem, while the reduced problem is equivalent to the original full problem in terms of commitment decisions. Our key insights lie on pre-screening the constraints based on the load distribution and considering the physical feasibility regions of multi-interval UC problem.  For the multi-step UC formulation, we overcome screening conservativeness by utilizing the multi-step ramping relationships, and can reliably screen out more constraints compared to current practice. Extensive simulations on both specific load samples and load regions validate the proposed technique can screen out more than 80\% constraints while preserving the feasibility of multi-interval UC problem.

%Unit commitment (UC) is a important task accomplished by independent system operator to determine the optimal schedule of unit status in daily operation of power systems. However, solving UC problems in a timely manner becomes challenging due to complexities and uncertainties of modern grids. In this paper, we develop a efficient and robust constraint screening routine based on chance-constrained optimization to eliminate redundant or inactive constraints from the original UC problem, so that the solution process on the reduced problem can be accelerate sufficiently and reliably. We conduct the offline screening for a given load region (sample-agnostic) to warm start the online screening for a load vector (sample-aware) belongs to the prescreened rregion. To overcome conservatives and redundancy of the existing constraint screening methods, we also utilize the cost information learned by machine learning (ML) models with load inputs to guide the screening procedure. Note that we take the load as the random variable with giving distribution in the screening model, then the perturbation of the load uncertainties on the constraint screening can be tackled with the chance-constrained optimization. (We verify the proposed method’s performance on a variety of UC setups and load situations, and the experimental results show that our method can screen out more operational constraints reliably.)
\end{abstract}

%\begin{IEEEkeywords}
%component, formatting, style, styling, insert
%\end{IEEEkeywords}

\section{Introduction}
\input{intro}

% \textcolor{red}{This paper is very related: Influence of Stochastic Dependence on Network
% Constraints Screening for Unit Commitment.
% }

% \textcolor{red}{From Line Roald SCOPF DRCC: While the assumption of a Gaussian distribution limits the
% applicability of the analytical reformulation from [7], [6], the
% analytical reformulation has some attractive properties. First,
% it is scalable to a large number of random variables, as the
% number of random variables does not influence the problem
% size or complexity of the OPF itself. Second, the solution
% is more transparent than a sample based solution since it
% is possible to trace the influence of each random variable
% through the analytical relations. }

\input{setup}
\input{experiment}

\input{Conclusion}
\bibliographystyle{IEEEtran}
\bibliography{bib}
\appendix
\input{appendix}
\end{document}

%% file: intro.tex
Obtaining accurate solutions for unit commitment in an efficient manner is crucial for ensuring reliable generation dispatch \cite{wang2008security}. For transmission grid operations, Unit commitment (UC) problems are typically formulated as mixed integer programming (MIP) problems involving both discrete variables for generator statuses and continuous variables for dispatch levels. In particular, for the multi-interval UC problems, temporal constraints, such as ramping constraints, are incorporated to regulate the capacity at which generators can adjust their output levels in response to dynamic changes of electricity demand and system conditions. However, due to the NP-hard nature of such nonconvex MIP problems, the solution process can be exceedingly time-consuming~\cite{bendotti2019complexity, Morales2013Tight}. Such computation complexity can be further increased by the existence of numerous network constraints such as line flow limits \cite{xavier2019transmission,guo2020fast}.

The need to accelerate the unit commitment (UC) solution process has prompted research into developing a surrogate UC model with fewer line flow limits while maintaining the solution equivalence of the resulting UC problem and the original UC problem. Such technique is backed up by the observation that only a subset of security constraints is binding in the real world systems. The process to identify the appropriate subset of line flow limits is termed \emph{constraint screening}~\cite{zhai2010fast,telgen1983identifying}. \cite{zhai2010fast} proposes to eliminate the constraints whose line flow cannot reach the boundary given load inputs and show that the feasible region will be the same as the original UC problem. 

However, for the standard screening model \cite{zhai2010fast, zhang2020data}, screening strategies can be conservative, while many redundant or inactive constraints are still kept rather than screened out. \cite{porras2021cost} and \cite{he2022enabling} propose the cost-driven screening models to utilize the operating cost constraint to limit the line flow value range further.
%, while the effect of the temporal constraints like ramping constraints is rarely considered in existing works. 
Yet most of the literature focus on single-step formulation of UC constraint screening, while ignoring the impact of temporal constraints such as generation ramping constraints on the feasible region of screening problems.  On the other hand, relaxed binary variables of generator commitment schedule in the standard screening model for a load sample also enlarge the line flow value range~\cite{roald2019implied}. The strong representation capability by machine learning (ML) models  \cite{Gao2023online,ding2020accelerating} can be utilized to predict the value of the binary variables, i.e., the decisions of generator states, which shows potential for further integrating the predictions to the screening model to handle the loose line flow range issue.

Another concern is the computation costs in standard constraint screening setup, which come from solving the optimization-based screening problem for each network constraint and  each electricity load instance. For the former, the ML models \cite{zhang2019data, pineda2020data,ardakani2018prediction, xavier2021learning,Cordero2022Warm-starting} directly use ML predictions to classify if a network constraint is redundant, while there is no guarantee the reduced UC is equivalent to the original one. For the latter, in fact, empirical evidence shows that the samples belonging to some typical load regions can have the same set of redundant constraints in their corresponding UC problems~\cite{roald2019implied}. It is then sufficient to implement screening on the load region rather than working on individual load sample \cite{ardakani2013identification, awadalla2022influence}.

To address such challenges, we develop one of the first multi-interval UC constraint screening models to narrow down the search space of line flow constraints reliably, leading to screening strategy with more efficient performance. Our key insights lie on integrating the temporal constraints, i.e., the ramping constraints to the standard screening model, and such approach can be applied for either given load region or individual load sample. The potential of utilizing the ML predictions of generator states to improve the screening efficiency is also explored. Specifically, we formulate a tractable linear programming problem for multi-interval UC, and prove that more inactive constraints can be eliminated without changing the feasible region of original UC problem. Moreover, our method can be flexibly integrated to screen constraints for either one specific load vector, or to achieve solver warm-start (offline constraint screening beforehand to get the reduced problem which can be later used for real-time operation problem) for a given load region. We term these two cases as \emph{sample-aware}~\cite{zhai2010fast} and \emph{sample-agnostic}~\cite{roald2019implied} constraint screening respectively. Specially, for the \emph{sample-aware} case, we further propose to make use of ML prediction on commitment schedule to better limit the search space of the constraint screening problem. In the sample-aware case on IEEE 39-bus system, our proposed method with ground truth generator states can achieve 95.3\% screening rate, achieving a boost compared to the standard constraint screening~\cite{zhai2010fast} (82\%). With partial predictions, the feasible rate of our method reach 84\% while the solution gap is only 0.32\%. In the sample-agnostic case on 118-bus, our procedure can also find the most compact form of UC problem after the more efficient screening process.

%Compared to the literature, our scheme can screen as many constraints as possible with theoretical guarantee while promise the computational efficiency. 
\begin{figure}[htbp]
\includegraphics[height=4cm, width=8.61cm]{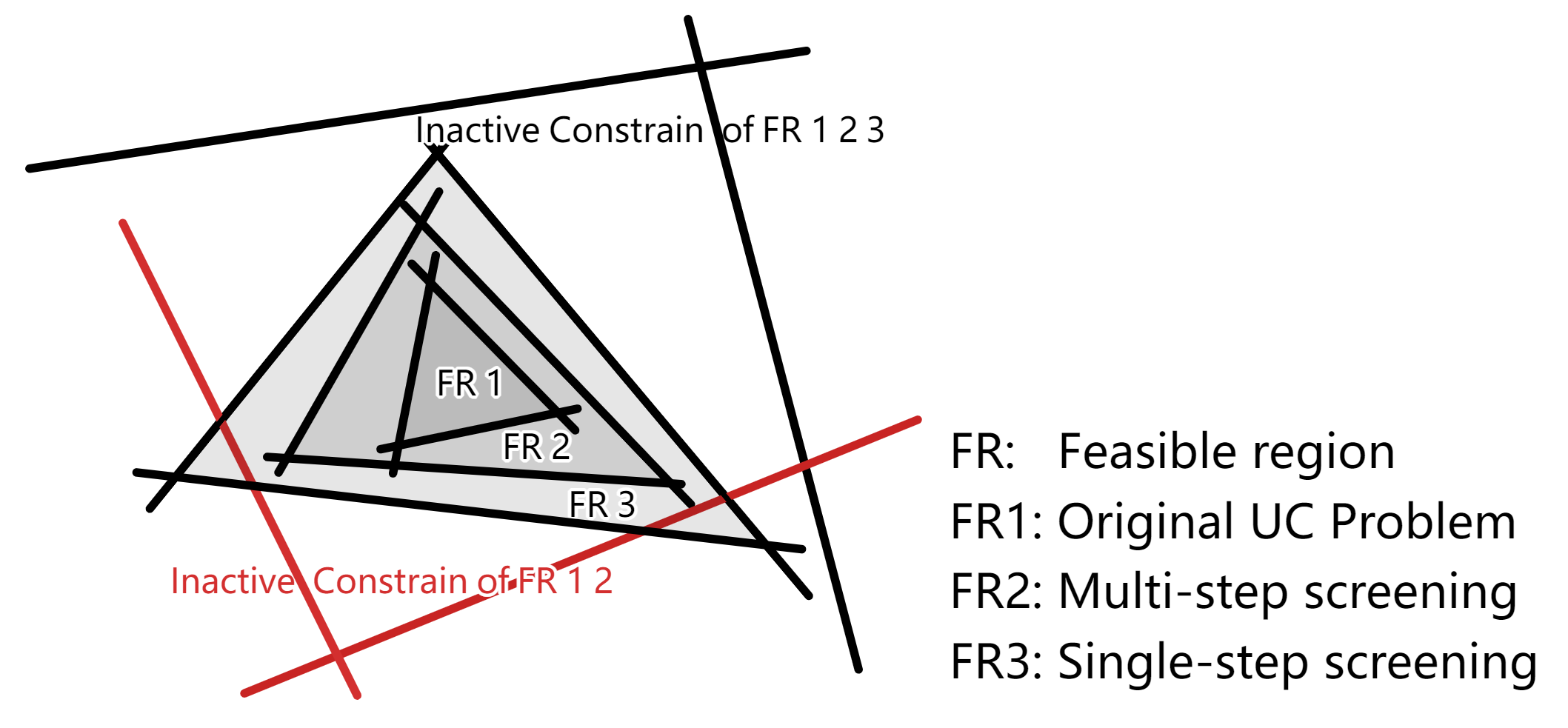}
\centering
\caption{Illustration of the inactive constraints corresponding to the different feasible region. By using our technique, it is possible to screen more constraints (line in red) compared to standard single timestep screening method.}\label{UC_FR}
% \todo{Jiayu: delete ''the`` in FR, FR1,... FR3; current-step to single-step (also change the name throughout the paper); use color coding for Fig1 and Fig2.}} \label{UC_FR}
\end{figure}

%% file: setup.tex
\section{Multi-Interval UC Problem Formulation}
In this paper, we assume the system operators need to decide both the ON/OFF statuses (the commitment schedule) as well as dispatch level for all generators. Herein we consider the day-ahead UC problem taking ramp constraints into consideration. For the $T$ timesteps UC problem with $n$ generators involved, the problem is formulated as
\begin{subequations}
% \vspace{0.5em}
\label{UC}
    \begin{align}
 \min _{\mathbf{u}, \mathbf{x}, \mathbf{f}}\quad   & \sum_{t=1}^{T}\sum_{i=1}^n c_i x_{i}(t) \label{UC:obj}\\
\text { s.t. } \quad  & u_i \underline{x}_i \leq x_i(t) \leq u_i \bar{x}_i, \quad \forall i, t \label{UC:gen}\\
& -\overline{\mathbf{f}} \leq \mathbf{K f(t)} \leq \overline{\mathbf{f}}, \quad \forall t \label{UC:flow}\\
& \mathbf{x(t)}+\mathbf{A} \mathbf{f(t)}=\boldsymbol{\ell}(t), \; \forall t \label{UC:balance}\\
& u_i(t) \in \{0, 1\}, \quad \forall i,t \label{UC:u}\\
& x_i(t)-x_i(t-1) \leq \text{R}^{up}_iu_{i}(t-1)\nonumber\\
& +\text{R}^{su}_i(u_i(t)-u_i(t-1)) + \bar{x}_i(1-u_i(t)) \; \forall i, t\label{UC:ramping_up}\\
& x_i(t-1) - x_i(t) \leq \text{R}^{dn}_iu_{i}(t)\nonumber\\
& +\text{R}^{sd}_i(u_i(t-1)-u_i(t)) + \bar{x}_i(1-u_i(t-1))\; \forall i, t. \label{UC:ramping_down}
\end{align}
\end{subequations}

In the UC problem, we optimize over the generator statuses $\mathbf{u}$, the generator dispatch $\mathbf{x}$ and the line power flow $\mathbf{f}$ to find the least-cost solutions with cost defined in the objective function \eqref{UC:obj}. $c_i$ denotes the cost coefficient. Constraint \eqref{UC:gen}, \eqref{UC:flow} and \eqref{UC:balance} denotes the generation bound, the flow bound and the nodal power balance respectively. Note that the power flows are modeled as a DC approximation, while the phase angles are absorbed into the fundamental flows $\mathbf{f}\in \mathbb{R}^{n-1}$~\cite{chen2022learning, bertsimas1997introduction}; $K$ and $\mathbf{A}$ map such fundamental flows to flow constraints and nodal power balance respectively. \eqref{UC:u} enforces the binary constraint of generator statuses, where $u_i=1$ indicates the generator is on. \eqref{UC:ramping_up} and \eqref{UC:ramping_up} are the ramping constraints to enforce limitations on the speed at which generators are able to modify their output levels in response to fluctuations in demand or system conditions. $R^{up}_i, R^{su}_i,R^{dn}_i,R^{sd}_i$ are the upward limits, downward, start-up and shut-down capacity.

For the power system with numerous generators and networks, \eqref{UC} can be a large-scale MILP problem, which is intractable to solve efficiently so as to satisfy the computation requirement of finding generation dispatch decisions. Meanwhile, there is a large number of inactive constraints, especially the inactive network constraints, giving the potential to simplify \eqref{UC} by screening out such inactive constraints. Then it is possible to solve a reduced UC problem with fewer engineering constraints.

% \todo{Describe why solving such a problem is hard, and why there is a need for constraint screening.}

%, differing from \cite{zhai2010fast}, we not only consider the constraints of single step but also that of previous $k$ steps. Furthermore, the ramping constraints are involved. 

\section{Multi-Interval Constraint Screening Model}
\label{sec:multi}
% \todo{This section contains many details, try to break up one section talking about the modeling, another section on how to solve the problem for sampel-aware and sample-agnostic case}
\subsection{Modeling of Inactive Constraints}
In this work, we follow the definition of inactive constraint firstly formulated by \cite{zhai2010fast}. One constraint can be treated as inactive if it has no influence on the feasible region of the multi-interval UC problem, and can thus be eliminated as illustrated in Fig. \ref{UC_FR}. For the constraint screening problem, it is thus of interest to correctly screen out as many inactive constraints as possible. And ultimately, we want to tighten the feasible region and make it close to the original UC problem, so constraints like those marked in red in Fig. \ref{UC_FR} can be screened out. 

To formulate the feasible region mathematically, we use $P$ to denote the feasible region defined by a constraint set $C^P$. $C^P$ is a subset of the original UC problem's constraint set. Then the actual feasible region of the original problem defined by \eqref{UC:gen}-\eqref{UC:ramping_down} can be naturally a subregion of $P$.
 
\emph{Definition 1:} A network constraint
$
\mathbf{K}_j \mathbf{f(t)} \leq \overline{\mathbf{f}}$ or $
\mathbf{K}_j \mathbf{f(t)} \geq -\overline{\mathbf{f}}$ is defined inactive to $P$ if and only if,
\begin{align}\label{UC_region}
P \supseteq P^{-\{j\}};
\end{align}
where $P^{-\{j\}}$ is the region defined by the constraint set $C^P$ eliminating $\mathbf{K}_j \mathbf{f(t)} \leq \overline{\mathbf{f}}_j$ or $
\mathbf{K}_j \mathbf{f(t)} \geq -\overline{\mathbf{f}}_j$, and we denote this set as $C^{P/j}$.

\subsection{Single-Step Constraint Screening}
To identify each inactive flow constraint, the standard optimization-based screening model\cite{zhai2010fast} tries to evaluate whether the line $j$ will be binding given the load input. In \eqref{screening2} we describe such formulation, which can be used to conduct the \emph{sample-aware constraint screening} for the $j$-th line at time step $k$.
% \vspace{-0.7em}
\begin{subequations}
\label{screening2}
    \begin{align}
\max_{\mathbf{u_k}, \mathbf{x_k}, \mathbf{f_k}} / \min _{\mathbf{u_k}, \mathbf{x_k}, \mathbf{f_k}}\quad   & \mathbf{K}_j \mathbf{f}(k)\\
\text { s.t. } \quad & u_i \underline{x}_i \leq x_i(k) \leq u_i \bar{x}_i, \quad i =1, ..., n \label{Screening2:gen}\\
&-\overline{\mathbf{f}}_{\mathcal{F}/j} \leq \mathbf{K}_{\mathcal{F}/j} \hat{\mathbf{f}}(k) \leq \overline{\mathbf{f}}_{\mathcal{F}/j}\label{Screening2:flow}\\
& \mathbf{x(k)}+\mathbf{A}\mathbf{f(k)}=\boldsymbol{\ell}(k) \label{Screening2:balance}\\
& 0\leq u_i(k) \leq 1. \quad i=1,...,n \label{Screening2:u};
\end{align}
\end{subequations}
where $\boldsymbol{\ell}$ is a known load vector for UC problem. Note that this formulation can be extended to eliminate some inactive constraints for multi-interval UC problem in \cite{zhai2010fast}, where $C^{P/j}={\eqref{Screening2:gen}-\eqref{Screening2:balance}}\cup\{u_i(t)\in\{0,1\}\}$. Note that \eqref{Screening2:u} relaxes $u$ as the continuous variables and thus \eqref{screening2} can be solved as a linear programming problem. However, the feasible region $P$ considered by this model ignore the impact of the ramping constraints, leading to a more relaxed screened region $P^{-\{j\}}$. This means more constraints can be identified as active for the screened region, while actually they are inactive for the feasible region of original UC problem, as shown in Fig. \ref{UC_FR}.

% For the case that the load region $\mathcal{L}$ is known, a \emph{sample-agnostic constraint screening problem} can be formulated for a group of operating scenarios, which can be given as follows,
% \begin{subequations}
% \vspace{-0.8em}
% \label{screening1}
%     \begin{align}
% \max_{\mathbf{u(t)}, \mathbf{x(t)}, \mathbf{f(t)}, \boldsymbol{\ell}(t)} / &\min _{\mathbf{u}(t), \mathbf{x}(t), \mathbf{f}(t), \boldsymbol{\ell}(t)}\quad   \mathbf{K}_j \mathbf{f(t)} \label{Screening:obj}\\
% &\text { s.t. } \quad \nonumber \eqref{Screening2:gen}-\eqref{Screening2:gen}\\
%   &\boldsymbol{\ell} \in \mathcal{L} \label{Screening2:u}
% \end{align}
% \end{subequations}
% where $\mathcal{F}/j$ denotes all remaining entries of vectors or matrix which excludes those corresponding to $f_j$.

\subsection{Multi-Step Constraint Screening}
\begin{figure}[h]
\includegraphics[height=3.1cm, width=6.8cm]{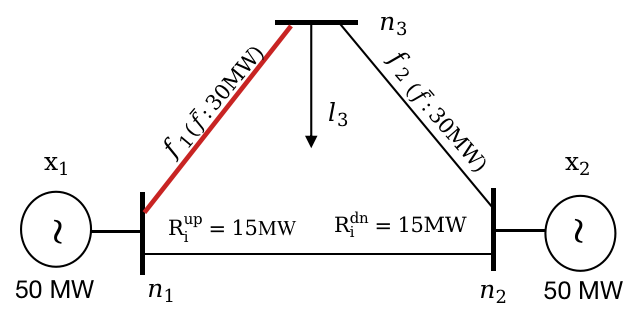}
\centering
\caption{Illustrative example of the three-node network.} \label{UC_Toy}
\end{figure}

In practice, multi-period UC is adopted for day-ahead scheduling. Yet in such multi-period problem, the existence of ramping constraints complicate both the solution procedure and also the analysis for constraint screening. Thus there are fewer previous literature working on this more realistic setup. Intuitively, ramping constraints can limit the generation range level and further limit the line flow regions.  Naturally, the ramping constraints can also influence the constraint screening process, and to illustrate this, consider the three-node network depicted in Fig. \ref{UC_Toy}.
\subsubsection{Influence of the ramping constraints on screening} Assume that we are screening $f_1(t+1)$ and considering the impact of a ramping-down constraint. Suppose we have load $l_3(t)=60\text{MW}$, $l_3(t+1)=35\text{MW}$, and we assume the upper bound of $f_1$ is $30$MW. Assume at timestep $t$, we know the dispatch solution $x_1(t)=30\text{MW}, x_2(t)=30\text{MW}$. Regarding the screening models maximizing the line flow of $f_1(t)$ we have: For the single-step screening, $x_1(t+1)=35\text{MW}, x_2(t+1)=0\text{MW}, f_1(t+1)=30\text{MW}$, then the upper bound of $f_1(t+1)$ is identified as active. For the multi-step screening, $x_1(t+1)=20\text{MW}, x_2(t+1)=15\text{MW}, f_1(t+1)=20\text{MW}$, then the upper bound of $f_1(t+1)$ is identified as inactive. 
% \todo{Need to describe the limit of line flow is 30 first.}
% \item Binding upward constraint: $l_3(1)=10\text{MW}, x_1(1)=25\text{MW}, x_2(1)=25\text{MW}$. When $l_3(2)=40\text{MW}$, and the screening models will maximize the $f_1(2)$. For the single-step screening, $x_1(2)=40\text{MW}, x_2(2)=0\text{MW}, f_1(2)=40\text{MW}$, then the upper bound of $f_1(2)$ is identified as active. For the multi-step screening, $x_1(2)=25\text{MW}, x_2(2)=15\text{MW}, f_1(2)=25\text{MW}$, then the upper bound of $f_1(2)$ is identified as inactive.

The above example illustrates that considering the ramping constraints in the screening can identify more inactive constraints, and such constraints can be eliminated from the original problem. Thus, we formulate the multi-step constraint screening model with ramping limits, and further prove that it can screen more constraints safely than the single-step screening.

\subsubsection{Model formulation}
 Without sacrificing generality, we first analyze the screening model \eqref{screening_con} for the upper bounds of the line limits, i.e., $\mathbf{K}_j \mathbf{f(t)} \leq \overline{\mathbf{f}}_j$, and the analysis for the lower bound can be achieved in the same manner.

\emph{Theorem 1:} Consider the following screening model for $j$-th line at time step $k$:
\begin{small}
\begin{subequations}
% \vspace{0.5em}
\label{screening_con}
    \begin{align}
 S_j^*(k) = \max_{\mathbf{u}_k, \mathbf{x}_k, \mathbf{f}_k} &\mathbf{K}_j \mathbf{f}(k)\label{UC_con:obj}\\
\text { s.t. } \quad & 
\mathbf{u}_k, \mathbf{x}_k, \mathbf{f}_k \in P^{-\{j\}}
\end{align}
\end{subequations}
\end{small}
A constraint $\mathbf{K}_j \mathbf{f}(k) \leq \overline{\mathbf{f}}_j$ is inactive to $P$ if and only if $S_j^*(k) \leq \overline{\mathbf{f}}_j$.

See proof in Appendix \ref{Proof_Theorem1}. Built upon on this theorem, we can define $C^{P/j}$ for the multi-step constraint screening model considering ramping constraints as follows,
\begin{small}
\begin{subnumcases}
~u_i(t) \underline{x}_i \leq x_i(t) \leq u_i(t) \bar{x}_i, \quad \forall i \in I, t\leq k \label{UC_con:gen}\\
-\overline{\mathbf{f}} \leq \mathbf{K}_{\mathcal{F}/j}\mathbf{f(t)} \leq \overline{\mathbf{f}}, \quad t \leq k \label{UC_con:flow}\\
\mathbf{x(t)}+\mathbf{A} \mathbf{f(t)}=\boldsymbol{\ell}(t), \quad t \leq k \label{UC_con:balance}\\
u_i(t) \in \{0, 1\}, \quad \forall i \in I,\; t\leq k \label{UC_con:u}\\
\scriptstyle x_i(t)-x_i(t-1) \leq \text{R}^{up}_iu_{i}(t-1)+\text{R}^{su}_i(u_i(t)-u_i(t-1)) \nonumber \\ 
\scriptstyle + \bar{x}_i(1-u_i(t)) \quad  
\forall i \in I, \; t\leq k, \label{UC_con:ramping_up}\\
\scriptstyle x_i(t-1) - x_i(t) \leq \text{R}^{dn}_iu_{i}(t)+\text{R}^{sd}_i(u_i(t-1)-u_i(t)) \nonumber \\
\scriptstyle + \bar{x}_i(1-u_i(t-1))\quad \forall i \in I, \;t\leq k \label{UC_con:ramping_down}
\end{subnumcases}
\end{small}
where $I$ denotes the bus index set. We denote the set \eqref{UC_con:gen}-\eqref{UC_con:ramping_down} as $C^{P/j}_m$ and corresponding feasible region as $P^{-\{j\}}_m$. By integrating $C^{P/j}_m$ to \eqref{screening_con}, we can get a multi-step screening model that can screen out inactive constraints reliably.

\subsection{Comparison of Single-Step and Multi-Step Screening}
One of our key insights come from including temporal relations \eqref{UC_con:ramping_up} and \eqref{UC_con:ramping_down} into the multi-interval constraint screening process. The constraint set of single-step screening \eqref{screening2} is a subset of $C^{P/j}_m$, then the feasible region of \eqref{screening2} covers that of $C^{P/j}_m$. This means that a constraint can be active for the feasible region of single-period problem \eqref{screening2}, while it is inactive for the feasible regions of $C^{P/j}_m$ and the original UC problem. Thus, it can be proved that the screening model derived by Theorem 1 can screen more inactive constraints for the original UC problem than the single-step screening reliably. 
% \todo{equ 6 is referenced before discussion}

\section{Improved model formulations for multi-step constraint screening}
Note that the multi-step screening model described in Section \ref{sec:multi} is still not tractable to solve due to the existence of binary variables, the cumbersome constraints and the regular fluctuation of load samples.  To improve the screening efficiency and develop realistic multi-step screening methods, in this section we further develop more efficient and reliable screening models.

\subsection{Sample-Aware Screening with Generator States}
Here \emph{sample-aware} means we implement constraint screening for a known load instance. As the binary variables make the basic screening model a MILP problem, we first consider relaxing the value of binary variables to $u \in [0,1]$. Then, to further lower the model complexity, we replace the nodal balance constraints \eqref{UC_con:balance} with the load balance constraints \eqref{UC_AW:nodal balance} and \eqref{UC_AW:balance}, and reduce the network constraints \eqref{UC_con:flow} to \eqref{UC_AW:flow}, as described in the following formulation \eqref{UC_AW}.

\emph{Corollary 1:} Consider the following multi-step screening model with the relaxation of the binary variables, part of nodal balance constraints and network constraints:
\begin{small}
\begin{subequations}
% \vspace{0.5em}
\label{UC_AW}
\begin{align}
 S_{aware_j}^*(k) = &\max_{\mathbf{u}_k, \mathbf{x}_k, \mathbf{f}_k} \;   \mathbf{K}_j \mathbf{f}(k)\label{UC_AW:obj}\\
\text { s.t. } \quad
  &\mathbf{\underline{x}} \leq \mathbf{x(t)} \leq \mathbf{\bar{x}}, \quad t\leq k, \label{UC_AW:gen}\\
& -\overline{\mathbf{f}}_{\mathcal{F}/j} \leq \mathbf{K}_{\mathcal{F}/j} \mathbf{f(k)} \leq \overline{\mathbf{f}}_{\mathcal{F}/j}, \quad  t=k,\label{UC_AW:flow}\\
 &\mathbf{x(k)}+\mathbf{A} \mathbf{f(t)}=\boldsymbol{\ell}(t), \quad t = k, \label{UC_AW:nodal balance}\\
& \sum{x(t)}=\sum{l(t)}, \quad t\le k, \label{UC_AW:balance}\\ 
& \nonumber 
 \eqref{UC_con:ramping_up}-\eqref{UC_con:ramping_down}.
\end{align}
\end{subequations}
\end{small}
If $ S_{aware_j}^*(k) \leq \overline{\mathbf{f}}_j$, then constraint $\mathbf{K}_j \mathbf{f}(k) \leq \overline{\mathbf{f}}_j$ is inactive to $P_m$, and thus is inactive to original UC problem.

% \todo{Xuan: Use Alignment similar to Corollary 1's form}

\begin{proof} With relaxing the binary variables, the nodal balance constraints, and the network constraints, the feasible region of \eqref{UC_AW} denoted as $P^{-\{j\}}_{aware}$ will cover $P^{-\{j\}}_m$, then naturally $S_j^*(k) \leq S_{aware_j}^*(k)$. Thus, $S_{aware_j}^*(k) \leq \overline{\mathbf{f}}_j$ implies that $S_j^*(k) \leq \overline{\mathbf{f}}_j$ and based on Theorem 1 the constraint $\mathbf{K}_j \mathbf{f}(k) \leq \overline{\mathbf{f}}_j$ will be inactive to $P_m$.
\end{proof}
Nevertheless, the relaxation of the binary variables can enlarge the screened feasible region, and thus line flow value range can be increased a lot. Then, more inactive constraints in the feasible region of the original problem are identified as active. Thus, it is of significance to limit the value of binary variables to narrow the screened feasible region. The recently emerging works on predicting the decisions of generator states, i.e., the binary variables using ML model \cite{Gao2023online,ding2020accelerating} give the potential that utilizing the ML predictions to aid the constraint screening. We denote the ground truth and the prediction of $\mathbf{u(t)}$ as $\mathbf{u}^*(t)$ and $\mathbf{\hat{u}}(t)$ respectively. The constraint set $C^{P/j}_m$ with $\hat{u}(t)$ is denoted as $C^{P/j}_{m,\hat{u}}$, and the corresponding feasible region is denoted as $P^{-\{j\}}_{m,\hat{u}}$.

% \emph{Lemma 1:}Assume that the ground truth $\hat{u}(t)$ of $u(t)$ in the original UC problem can be totally and accurately predicted. When fixing the value of $u(t)$ to $\hat{u}(t)$ for constraint set $S_1$ and $S_2$, and denoting the new set as $S^u_1$ and $S^u_2$. If $S_1$ is a$S^{P/j}$ corresponding to $S_2 = S^P$, then obivously $S^u_1$ is also $S^{P/j}$ corresponding to $S^u_2 = S^P$.

\emph{Corollary 2:} When $\mathbf{\hat{u}}(t)=\mathbf{u}^*(t)$ holds for all time steps, consider the following screening model.
\begin{small}
\begin{subequations}
% \vspace{0.5em}
\label{UC_GUG}
    \begin{align}
 S_{truth_j}^*(k) = & \max_{\mathbf{x}_k, \mathbf{f}_k}\mathbf{K}_j \mathbf{f}(k)\label{UC_GUG:obj}\\
\text { s.t. } \quad
& \hat{u}_i(t) \underline{x}_i \leq x_i(t) \leq \hat{u}_i(t) \bar{x}_i,  \forall i \in I, t\leq k, \label{UC_con2:gen} \\
& \nonumber \eqref{UC_con:ramping_up}-\eqref{UC_con:ramping_down}, \eqref{UC_AW:flow}-\eqref{UC_AW:balance}.
\end{align}
\end{subequations}
\end{small}

If $ S_{truth_j}^*(k) \leq \overline{\mathbf{f}}_j$, then the constraint $\mathbf{K}_j \mathbf{f}(k) \leq \overline{\mathbf{f}}_j$ is inactive to $P_{m,\hat{u}}$, and thus is inactive to original UC problem with $\hat{u}(t)$.

\begin{proof}
     $C^{P/j}_{m,\hat{u}}$ is obviously still $C^{P/j}$ corresponding to $C^{P}_m$ with $\hat{u}(t)$. Let $P^{-\{j\}}=P^{-\{j\}}_{m,\hat{u}}$ in \eqref{screening_con}, then if $S_j^*(k) \leq \overline{\mathbf{f}}_j$, according to Theorem 1  $\mathbf{K}_j \mathbf{f}(k) \leq \overline{\mathbf{f}}_j$ is inactive to $P_{m,\hat{u}}$, and thus is inactive to original UC problem with $\hat{u}(t)$. Further, as the feasible region of \eqref{UC_GUG} covers $P_{m,\hat{u}}$, then we have $S_j^*(k) \leq S_{truth_j}^*(k)$. Thus, $ S_{truth_j}^*(k) \leq \overline{\mathbf{f}}_j$ can imply $S_j^*(k) \leq \overline{\mathbf{f}}_j$ so as to identify the inactive constraint. 
\end{proof} 

 Actually, the ML predictions are not always reliable, which means $\mathbf{\hat{u}}(t)=\mathbf{u}^*(t)$ may not hold. Then, the predictions $\mathbf{\hat{u}}_k=\{\mathbf{\hat{u}}(1),\mathbf{\hat{u}}(2),...,\mathbf{\hat{u}(k)}\}$ can cause the infeasible cases for both the original UC model and the screening model. Thus, we consider only replacing a part of binary variables with accurate ML predictions~\footnote{This is achievable for some critical nodes or timesteps by learning the UC binary decisions of a reduced problem.}, such as for specified time steps $t \in T_{pre}$, we give the predictions $\hat{u}_i(t)={u}^*_i(t), \forall i \in I$ as shown in \eqref{UC_GUP:u}. As long as the inserted predictions are accurate, the feasible solutions $\mathbf{u}^{orig}_{T}=\mathbf{u}^{*}$ for \eqref{UC} and $\mathbf{u}^{scre}_{k}$ for \eqref{UC_GUP} can be found. Specifically, with the mixed decision vectors $\mathbf{u}_T$ to solve the original UC problem \eqref{UC} and $\mathbf{u}_k$ to solve the screening problem \eqref{UC_GUP}, we can get the corresponding solutions involving the value of generator states and the generations, where the rest solutions of ${u}_i(t), \forall i \in I,t \notin T_{pre}$ are determined. Let $\mathbf{u}^{orig}_{k} = \{\mathbf{u}^{orig}(1),\mathbf{u}^{orig}(2),...,\mathbf{u}^{orig}(k)\}$, the case that $\mathbf{u}^{orig}_{k} \neq \mathbf{u}^{scre}_{k}$ may occur due to the different objective functions. The next Corollary states that under such case, our screening model can still promise effectiveness to identify the inactive constraints correctly.

 % \todo{What is with solving the optimization...? And what does c and s mean in the superscript? Why they are feasible? Explain more.} With solving the optimization problems, we may get the feasible solutions $\mathbf{u^{orig}_{f}}(k)$ and $\mathbf{u^{scre}_{f}}(k)$ for the original UC model and the screening model respectively. 

\emph{Corollary 3:} Consider the following screening model with partial ML predictions:
\begin{small}
\begin{subequations}
% \vspace{0.5em}
\label{UC_GUP}
    \begin{align}
 S_{partial_j}^*(k) = & \max_{\mathbf{x}_k, \mathbf{f}_k, \mathbf{u}_k}\mathbf{K}_j \mathbf{f}(k)\label{UC_GUG:obj}\\
\text { s.t. } 
 \quad  & {u}_i(t) = \hat{u}_i(t),  i \in I, \quad t \in T_{pre}, 
 \label{UC_GUP:u} \\
 & {u}_i(t) \underline{x}_i \leq x_i(t) \leq u_i(t) \bar{x}_i,  \forall i \in I, \quad t\leq k, \label{UC_GUP:gen} 
 \\
& \nonumber \eqref{UC_con:ramping_up}-\eqref{UC_GUP}, \eqref{UC_AW:flow}-\eqref{UC_AW:balance}.
\end{align}
\end{subequations}
\end{small}

If $ S_{partial_j}^*(k) \leq \overline{\mathbf{f}}_j$, the constraint $\mathbf{K}_j \mathbf{f}(k) \leq \overline{\mathbf{f}}_j$ is inactive to $P_{m,u^c_{f}(k)}$, and thus is inactive to the original UC problem with $\mathbf{u}^{orig}_k$.

% When $\mathbf{u^c_{f}}(k)$ and $\mathbf{u^s_{f}}(k)$ exist, 

\begin{proof}
If $S_{partial_j}^*(k) \leq \overline{\mathbf{f}}_j$, then the value of $S_{partial_j}(k)$ corresponding to $\mathbf{u}^{orig}_k$ will always satisfy $S_{partial_j}^*(k) \leq S_{partial_j}^*(k) \leq \overline{\mathbf{f}}_j$. Thus, the constraint $\mathbf{K}_j \mathbf{f}(k) \leq \overline{\mathbf{f}}_j$ is inactive to the original UC problem with $\mathbf{u}^{orig}_k$.
\end{proof}

Therefore, with the accurate and partial predictions of $u(t)$, we can still eliminate the inactive constraints identified by \eqref{UC_GUP} safely. Note that Corollary 2 and Corollary 3 can also be extended easily to other screening models with the different definitions of $P$, e.g., the cost-driven screening model proposed by \cite{porras2021cost}.
% \todo{Yize: Just describe the differences of our C2 and C3 to others at this place; cite others' work}
%With the partial predictions of $u(t)$, if there is a feasible solution, using the following model to conduct constraint screen will not misidentify the active constraint as inactive for the original UC problem with the same solution.

\subsection{Sample-Agnostic Screening with Typical Load Region}
In practice, the group of active constraints tend to remain the same across different realizations of load samples for given power grid, motivating methods that identify a collection of active constraints with varying load data. Our framework can also be extended to such setup, and we term it as \emph{sample-agnostic screening}. We define a load region $\mathcal{L}$, and we assume that all the possibly upcoming load samples $\boldsymbol{\ell} \in \mathcal{L}$, while in the screening model the load sample is transformed to $\boldsymbol{\ell}^r$.

% \todo{give mathematical definition of load samples, and denote  that covers all the possibly upcoming load samples.}

\emph{Corollary 4:} Consider the following multi-step screening model for a specified load region:
\begin{small}
\begin{subequations}
% \vspace{0.5em}
\label{UC_GR}
    \begin{align}
 S_{R_j}^*(k) = & \max_{\mathbf{u}_k, \mathbf{x}_k, \mathbf{f}_k,\mathbf{l}^r_k}\mathbf{K}_j \mathbf{f}(k)\label{UC_con2:obj}\\
\text { s.t. } \quad
 & \mathbf{\underline{x}} \leq \mathbf{x(t)} \leq \mathbf{\bar{x}}, \quad t\leq k, \\
& -\overline{\mathbf{f}}_{\mathcal{F}/j} \leq \mathbf{K}_{\mathcal{F}/j} \mathbf{f(t)} \leq \overline{\mathbf{f}}_{\mathcal{F}/j}, \quad  t=k, \label{UC_con3:flow}\\
& \mathbf{x(t)}+\mathbf{A} \mathbf{f(t)}=\boldsymbol{\ell}^{r}(t), \quad t = k, \label{UC_con3:balance}\\
& \sum{x(t)}=\sum{l^r(t)}, \quad  t\le k, \\
& \boldsymbol{\ell}^{r} \in \mathcal{L} \label{UC_con3:l},\\
& \eqref{UC_con:ramping_up}-\eqref{UC_con:ramping_down}.\nonumber
\end{align}
\end{subequations}
\end{small}
If $ S_{region_j}^*(k) \leq \overline{\mathbf{f}}_j$, then constraint $\mathbf{K}_j \mathbf{f}(k) \leq \overline{\mathbf{f}}_j$ is inactive to $P_m$, and thus is inactive to original UC problem.

\begin{proof}
With relaxing the binary variables and load value, the feasible region of \eqref{UC_GR} denoted will cover $P^{-\{j\}}_m$, then naturally $S_j^*(k) \leq S_{region_j}^*(k)$. Thus, $S_{region_j}^*(k) \leq \overline{\mathbf{f}}_j$ implies that $S_j^*(k) \leq \overline{\mathbf{f}}_j$ and based on Theorem 1 the constraint $\mathbf{K}_j \mathbf{f}(k) \leq \overline{\mathbf{f}}_j$ will be inactive to $P_m$. 
\end{proof}

% Similar relaxation can be done for the single-step screening, which means the constraint \eqref{UC_con3:l} can be integrated to \eqref{screening2}.

\begin{figure}[b]
\vspace{-1.5em}
\includegraphics[height=5.4cm, width=8cm]{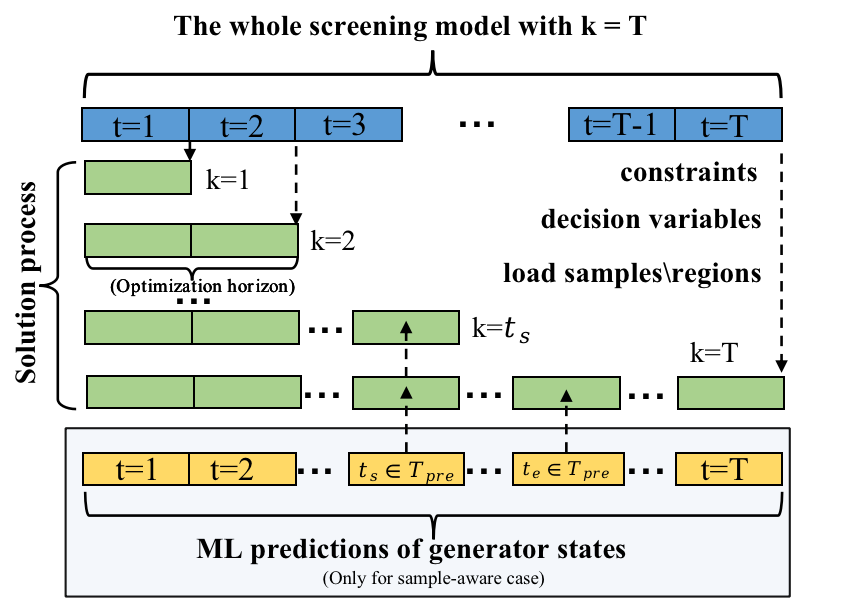}
\centering
\vspace{-0em}
\caption{The schematic of the solution process of the improved models.} \label{UC_solution}
\vspace{-1.5em}
\end{figure}

% can be applied for the short-term (day-ahead), medium-term (week-ahead) and long-term (over several months to year) UC problem, and 

\section{Solution for the improved models}
As we describe in the previous sections,  proposed screening model can be treated as a pre-processing step that needs to be solved before the optimization process of UC problem. In this paper, we will solve the optimization models derived by Corollary 2-Corollary 4, and the solution process of the improved models is shown in Fig. \ref{UC_solution}.  This process is conducted for each time step $t=1,2,...,T$, and the optimization horizon is equal to the current time step $k$, making the solution process a rolling problem.
\subsection{Solution for Sample-Aware Screening}
For the sample-aware case, the solution process involves the prediction part as illustrated in Fig. \ref{UC_solution} by the yellow blocks which represent the predictions of generator states $\mathbf{\hat{u}}(t)$. Regarding the solution process with ground truth, we get the ground truth by solving the original multi-interval UC problem and record the generator states $\mathbf{{u}}^*(t)$. Then, we let $\mathbf{\hat{u}}(t)=\mathbf{{u}}^*(t), t \leq k$ and solve the model \eqref{UC_GUG} with the optimization horizon equal to $k$.

Regarding the solution process with partial predictions, here we use k-Nearest Neighbor (KNN) method to get the predictions due to its simplicity, interpretability, and flexibility \cite{pineda2020data}. KNN calculates the distance between the incoming load sample and the historical load samples, and then predicts the ON/OFF states of generator according to the plurality vote of its closet K neighbors. Once the prediction $\mathbf{\hat{u}}(t)$ is obtained, we have $\mathbf{\hat{u}}(t)=\mathbf{{u}}^*(t), t \in T_{pre}$ and solve the model \eqref{UC_GUP} with the optimization horizon equal to $k$.

\subsection{Solution for Sample-Agnostic Screening}
In this case,  without the prediction part, we solve the screening model \eqref{UC_GR} for a load region which needs to be formulated firstly. Herein, we define a load range denoted as $r$, centered around the nominal load $\overline{\boldsymbol{\ell}}_t$, then the load region $\mathcal{L}$ can be given as follows,
\begin{equation}
(1-r)\overline{\boldsymbol{\ell}}(t)\leq \boldsymbol{\ell}^r(t)\leq(1+r)\overline{\boldsymbol{\ell}}(t).
\label{UC_solu:load range}
\end{equation}
Then, we can replace \eqref{UC_con3:l} with \eqref{UC_solu:load range} and solve the sample-agnostic screening model \eqref{UC_GR} with the optimization horizon equal to $k$.

Note that the models for the lower bound of line limits will be solved identically. Once the screening models are solved, we can get the binding situation for each line limit at each time step of the original multi-interval UC problem. Then, by eliminating the constraints identified as inactive, we can get to the reduced multi-interval UC problem. 
% \subsection{Solution for Sample-Agnostic Screening}
% In this case, we collect the varying data and formulate the load region. Then the screening model will be solved in a rolling way. For the solution of $k$ step, 

%% file: experiment.tex
\section{Case study}

In this Section, we evaluate the performance of the proposed multi-step constraint screening models, and compare the number of remaining constraints and solution time with those of the original problem and the single-step models. We demonstrate the proposed multi-step screening procedure is both effective and reliable.

% For \emph{sample-agnostic} case, we examine the performance of the proposed models on different load regions.

\subsection{Simulation Setup}

In our experiment, we conduct numerical simulations on IEEE 39-bus and IEEE 118-bus power systems to evaluate the effectiveness of our proposed method on both small-scale and larger-scale system. Specifically, we test $r$ taking values of 20\%, 50\%, 80\%, and the nominal load profiles are coming from real 24-hour load dataset.

For the sample-aware screening with prediction, we take KNN as the prediction model. To generate samples for training and validating KNN model, we use the uniform distribution to get random ${\boldsymbol{\ell}_t}$ with $r=50\%$ and then solve (\ref{UC}) for all generated loads. The states of each generator for 24 hours are recorded. 2,500 samples of 39-bus system are solved for KNN training. Moreover, when evaluating the screening performance of the multi-step and single-step method, we use the same validation data and consider 50 samples for each validation case.

To achieve the screening with partial predictions, we first evaluate the KNN model's performance in Table \ref{table 1}. Based on the prediction accuracy and the feasibility of the original problem, we find K$=$5 and Interval$=$4 to be an appropriate choice so as to get as many feasible solutions as possible. 

% KNN prediction can make mistakes and thus causing infeasible cases. To handle this issue, we choose to insert prediction with a fixed time interval, which can promise higher likelihood to get the feasible solutions. 

\begin{table}[htbp]
\vspace{0em}
\centering
\caption{Parameter Selection of with \\ Available ML Model's Partial Prediction } \label{table 1}
\vspace{-0em}
\setlength{\tabcolsep}{3mm}{
\begin{threeparttable}{
\begin{tabular}{c|ccccc}
\hline
\textbf{K}\textsuperscript{1} & 2 & \textbf{5} & 10 & 20 & 40\\ \hline
Error\textsuperscript{2} & 6.44\% & \textbf{5.10\%} & 4.87\% & 4.70\% & 4.65\% \\
Feasible & 49\% & \textbf{84\%} & 68\% & 77\% & 79\% \\ \hline
\textbf{Interval}\textsuperscript{3} & 2 & 3 & \textbf{4} & 5 & 6 \\ \hline
Feasible & 45\% & 78\% & \textbf{84\%} & 85\% & 45\% \\ \hline
\end{tabular}}
 \begin{tablenotes}
        \footnotesize
        \item 1. The number of nearest neighbours in KNN (Fixed Interval=4)
        \item 2. The classcification error of generator states
        \item 3. The interval at which the knn prediction is inserted (Fixed K=5)
      \end{tablenotes}
  \end{threeparttable}
}
\vspace{-1.5em}
\end{table}

\begin{table}[htbp]
\vspace{0em}
\centering
\caption{Number of Remaining \\Constraints after Sample-Aware Screening} \label{table2}
\vspace{-0em}
\setlength{\tabcolsep}{2.1mm}{
\begin{threeparttable}{
\begin{tabular}{c|ccc|c|c} 
\hline
Load Range & 20\% & 50\% & 80\% & Average & Decline\\ \hline
Line Constraint\textsuperscript{1} & 483 & 478 & 472 & 478 & -78.36\% \\
Line Constraint\textsuperscript{2} & 419 & 417 & 412 & 416 & -81.16\% \\ \hline
\end{tabular}}
 \begin{tablenotes}
        \footnotesize
        \item 1. Remaining constraints after single-step screening
        \item 2. Remaining constraints after multi-step screening
      \end{tablenotes}
  \end{threeparttable}
}
\vspace{-1.5em}
\end{table}

\begin{table}[htbp]
\vspace{-0em}
\centering
\caption{CPU Time for Solving \\Original Problem after Sample-Aware Screening} \label{table3}
\vspace{-0em}
\setlength{\tabcolsep}{2.1mm}{
\begin{tabular}{c|ccc|c|c} 
\hline
Load Range & 20\% & 50\% & 80\% & Average & Speed-up\\ \hline
Original Problem & 2.04 & 2.20 & 1.95 & 2.06 & - \\
Single-step& 1.86 & 1.92 & 1.80 & 1.86 & 1.107 \\ 
Multi-step & 1.84 & 1.88 & 1.79 & 1.84 & 1.124 \\ \hline
\end{tabular}
}
\vspace{-1em}
\end{table}

\begin{table}[htbp]
% \vspace{-1em}
\centering
\caption{Result of Sample \\Aware Screening with Partial Prediction} \label{table4}
\vspace{-0em}
\setlength{\tabcolsep}{2.3mm}{
\begin{threeparttable}{
\begin{tabular}{c|c|c|c}
\hline
Method & Line Constraints & CPU Time & Speed-up\\ \hline
Original Problem & 2280 & 2.20 & - \\ \hline
Single-step & 478 & 1.98 & 1.108  \\
Multi-step & 416 & 1.92 & 1.143 \\  \hline
Single-step\textsuperscript{*} & 414 & 1.17 & 1.873 \\ 
Multi-step\textsuperscript{*} & 287 & 1.06 & 2.065 \\
\hline
\end{tabular}}
 \begin{tablenotes}
        \footnotesize
        \item * Insert KNN prediction every 4 hours
      \end{tablenotes}
  \end{threeparttable}
}
\vspace{-1.5em}
\end{table}

\begin{table}[htbp]
% \vspace{-1.8em}
\centering
\caption{Number of Remaining \\Constraints after Sample Agnostic Screening} \label{table5}
% \vspace{-0.9em}
\setlength{\tabcolsep}{2mm}{\begin{tabular}{c|ccc|ccc}
\hline
\multirow{2}*{\diagbox{Scale}{Region}} & \multicolumn{3}{c|}{Multi-step Method} & \multicolumn{3}{c}{Single-step Method} \\ \cline{2-7}
 ~ &  20\% & 50\% & 80\%  & 
 20\% & 50\% & 80\% \\ \hline
39-bus  & 494 & 623 & 716 & 558 & 691 & 767\\
118-bus & 2283 & 2655 & 2978 & 2393 & 2744 & 3057
\\ \hline
\end{tabular}}
\vspace{-1.5em}
\end{table}

\subsection{Simulation Results}
\begin{figure}[h]
\vspace{-1.em} 
\includegraphics[height=4.5cm]{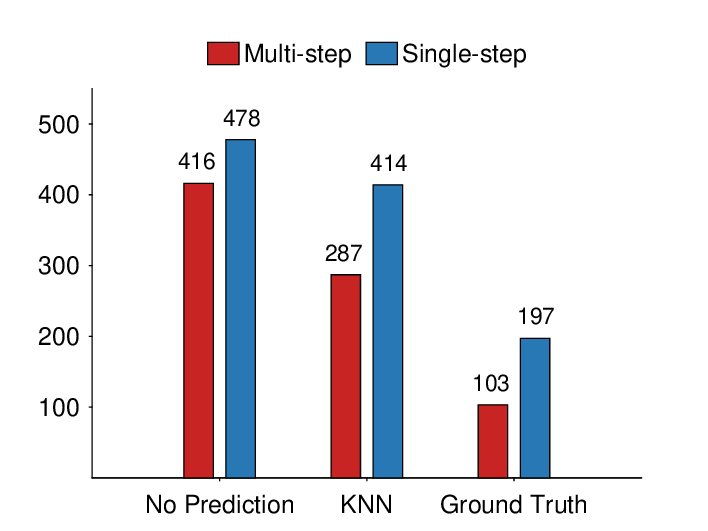}
\centering
\caption{The Comparison among Sample Aware \\Screening under Different Settings}\label{fig4}

\end{figure}

\emph{ Sample-aware screening tasks}: We consider three scenarios:  i) without generator statuses predictions, ii) with partial predictions, and iii) with ground truth are considered to verify the effectiveness of the proposed method. When there is no generator statuses predictions, our multi-step method reduces 81.16\% of line limits as shown in Table.\ref{table2}. With partial predictions, the number of remaining constraints of multi-step is only 69\% of single-step method, and the CPU time solving the original problem is 1.103 times faster than single-step method according to Table \ref{table4}. It can be seen that multi-step method can always eliminate more inactive constraints, and thus taking less time to solve the reduced UC problem. Besides, the infeasible rate of the original problem is only 16\%, and the gap of objective value between the reduced and the original UC problem is only 0.32\%. Note that the bias between the solution of binary variables obtained by the screening model and the UC problem does not cause the infeasible case due to the line limits misidentification, which verifies Corollary 3. According to Fig. \ref{fig4}, the case with the ground truth can eliminate the most inactive constraints, while with the partial predictions the remaining constraints can still be sufficiently reduced compared to the case without prediction.

\begin{figure}[h]
\vspace{-2.5em}
\includegraphics[height=3.5cm]{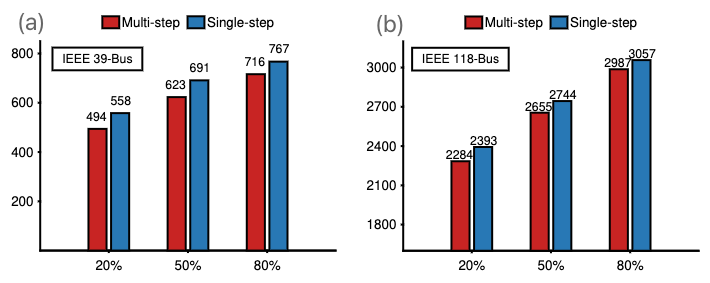}
\centering
\caption{The Number of Remaining Constraints \\after Sample Agnostic Screening in (a). IEEE 39-Bus and (b). 118-Bus systems.}\label{fig6}
\vspace{0em}
\end{figure}

\emph{ Sample-agnostic screening tasks}: In this setting, the cases both on 39-bus and 118-bus systems with 20\%, 50\%, and 80\% load variations are considered to verify the scalability of the proposed method. As shown in Fig. \ref{fig6}, the number of remaining constraints decrease by 60 and 100 for 39-bus and 118-bus system over three load ranges respectively. It can be seen that with larger load region, the number of remaining constraints increases. This may be due to the increasing patterns of inactive constraints with wider load variation range, i.e., the percentage of the inactive constraints decreases when widening the load variation. Results above validate that the proposed multi-step screening models can help boost screening performance accurately in both sample-aware and sample-agnostic settings. This demonstrates both the possibility and the necessity of including the practical multi-step ramping information into the screening problem.

%\footnotesize
%    \item * KNN: Parital Prediction Model (interval=4)
%    \item * Ground Truth: simulating 100\% accurate %prediction to explore the limitation (interval=1)
%\begin{figure}[h]
%\vspace{-2.5em}
%\includegraphics[height=4.5cm]{IEEE_39.eps}
%\centering
%\caption{The Number of Remaining Constraints \\after Sample Agnostic Screening in IEEE 39-Bus}\label{fig6}
%\vspace{0em}
%\end{figure}

%\begin{figure}[H]
%\vspace{-2em}
%\includegraphics[height=4.5cm]{IEEE_118.eps}
%\centering
%\caption{The Number of Remaining Constraints \\after Sample Agnostic Screening in IEEE 118-Bus}\label{fig5}
%\vspace{0em}
%\end{figure}

%% file: Conclusion.tex
\section{Conclusion}
In this paper, we develop a unified framework which includes the basic and the improved multi-step screening models with theoretical guarantees. The temporal constraints and the ML predictions of generator states are incorporated into the screening models, tightening the searching space of constraint screening sufficiently. The flexibility of screening for a load region further improves the efficiency of the proposed models. %Compared to the single-step screening model, our method can overcome screening conservativeness and thus eliminating more inactive constraints reliably. 
As we only consider the influence on the searching space from the explicit constraints of  UC problem, in the future work we
would like to seek theoretical understandings about the implicit decision spaces impacting the screening and final UC solution results.

%% file: appendix.tex
\subsection{Proof of Theorem 1} \label{Proof_Theorem1}
Our proof is inspired by the proof given in \cite{zhai2010fast}, but since the definition of $P$ in this paper differs from that of \cite{zhai2010fast}, so we rewrite the proof and fit it to our setup here.

\begin{proof}
\textbf{1) Sufficiency} If $ S_j^*(k) \leq \overline{\mathbf{f}}_j$, then any feasible solution of \eqref{screening_con} in $P^{-\{j\}}$, the following equation holds,
\begin{small}
% \vspace{0.5em}
\begin{align} \label{UC_Suffi}
\mathbf{K}_j \mathbf{f}(k) \leq S_j^*(k) \leq \overline{\mathbf{f}}_j 
\end{align}
\end{small}
Then, \eqref{UC_Suffi} means any feasible solution to $P^{-\{j\}}$ is also a feasible solution to $P$, i.e., 
\begin{align}\label{UC_region2}
P \supseteq P^{-\{j\}}
\end{align}
Thus, the constraint $\mathbf{K}_j \mathbf{f}(k) \leq \overline{\mathbf{f}}_j$ is inactive to $P$ according to Definition 1, and this can imply that the constraint is also inactive to the original UC problem as illustrated in Fig. \ref{UC_FR}.

\textbf{2) Necessity} If the constraint $\mathbf{K}_j \mathbf{f}(k) \leq \overline{\mathbf{f}}_j$ is inactive, \eqref{UC_region} holds. The inactive constraint implies that $\mathbf{K}_j \mathbf{f}(k) \leq \overline{\mathbf{f}}_j$ is always satisfied for any feasible solution in $P$, which means $\mathbf{K}_j \mathbf{f}(k) \leq \overline{\mathbf{f}}_j$ always holds for the feasible solution in $P^{-\{j\}}$. Thus, $S_j^*(k)\leq \overline{\mathbf{f}}_j$ is satisfied.
\end{proof}

\begin{figure}[h]
\vspace{-1.em}
\includegraphics[height=6cm]{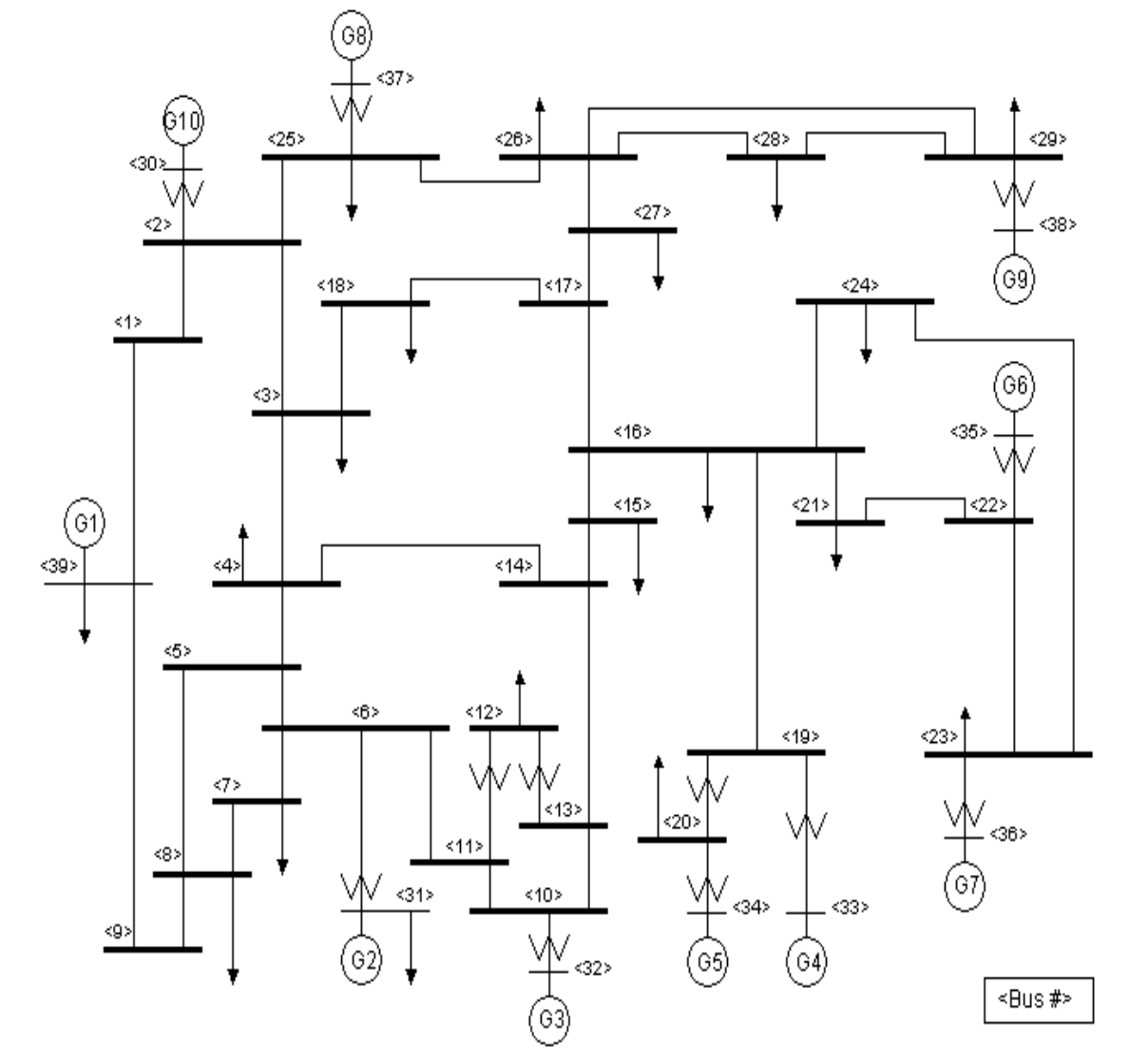}
\centering
\caption{IEEE 39-Bus System}\label{fig7}
\vspace{-1em}
\end{figure}

\subsection{A Brief Illustration to Constraint Screening}

To better illustrate the result of constraint screening, we choose IEEE 39-bus as an example in Fig.\ref{fig7}. In the original UC problem, it has 46 power lines, and the power flow of each line has lower bound and upper bound at each time step. Actually, not all such constraints are active in the original problem, motivating the research of constraint screening.

We take a specific sample from our simulation for demonstrating the case, where the real binding lower bound of the power line for current step are Line 24[15-16]\textsuperscript{*}, 36[22-35], and the active constraints in our multi-step method are Line 3[2-25], 24[15-16], 26[16-19], 34[21-22], 36[22-35], 38[23-36]. Meanwhile, single-step method remains ten more inactive constraints than our method. By eliminating theses inactive constraints, we can take less time to solve the reduced UC problem.

\textsuperscript{*} Line $x$[$i$-$j$]: the $x$-th line is from the $i$-th bus to the $j$-th bus in IEEE 39-bus system.